\documentclass[]{acmart} 

\setcopyright{acmlicensed}
\copyrightyear{2026}
\acmYear{2026}

\begin{document}

\title{Challenges in Working Towards Patient Engagement in
Developing Technology Prototypes}

\author{Fateme Rajabiyazdi}
\email{fateme.rajabiyazdi@ucalgary.ca}

\affiliation{%
  \institution{University of Calgary}
  \city{Calgary}
  \state{Alberta}
  \country{Canada}
}

\author{Julie Babione}
\affiliation{%
  \institution{University of Calgary}
  \city{Calgary}
  \country{Canada}}
\email{julie.babione@ucalgary.ca}

\author{Doreen M. Rabi}
\affiliation{%
  \institution{University of Calgary}
  \city{Calgary}
  \country{Canada}
}

\author{Foroozan Daneshzand}
\affiliation{%
 \institution{Simon Fraser University}
 \city{Burnaby}
 \state{British Columbia}
 \country{Canada}}
 \email{fdaneshz@sfu.ca}

\author{Sheelagh Carpendale}
\affiliation{%
  \institution{Simon Fraser University}
  \city{Burnaby}
  \country{Canada}}
\email{sheelagh@sfu.ca}

\renewcommand{\shortauthors}{Rajabiyazdi et al.}
\acmCodeLink{https://github.com/borisveytsman/acmart}
\acmDataLink{htps://zenodo.org/link}
\acmContributions{BT and GKMT designed the study; LT, VB, and AP
  conducted the experiments, BR, HC, CP and JS analyzed the results,
  JPK developed analytical predictions, all authors participated in
  writing the manuscript.}

\maketitle

\section{Introduction}

Creating supportive technologies for people living with multiple chronic conditions is extremely challenging. These patients are often faced with substantial visible and invisible treatment work as well as their everyday responsibilities, including coordinating across providers, tracking information, and repeating communication in emotionally charged contexts. In the Cumulative Complexity Model (CuCoM)\cite{shippee2012cumulative}, the balance between patient workload and patient capacity shapes what patients can realistically take on, including whether a digital tool can be adopted and sustained. In this paper, we report engagement lessons from implementing MyCareCompass, a patient-facing digital health intervention (DHI) intended to support day-to-day self-management for people living with multiple chronic conditions. We define engagement as patient uptake and sustained use during a two-month pilot study of our platform, drawing on usage analytics and follow-up feedback, and distill three implementation lessons for designing for engagement in complex chronic care. These insights align with the workshop's aim to bridge disparate conceptualizations of engagement across HCI, psychology, and implementation science.


\section{Human-Centered Design Process}

We used a human-centered design (HCD) approach to understand patient and clinician experiences and translated those insights into design artefacts that guided design and iterative development \cite{giacomin2014human,altman2018design,bazzano2017designing}. We gathered insights through interviews and focus groups, synthesized them using affinity diagramming and journey mapping to surface workload and capacity tensions, and identified where a tool might reduce workload through automation and increase capacity through supportive practices and integration. To keep design grounded in real contexts of use, we created personas and storyboards that combined patient stories into likely use case scenarios, and used these artefacts to guide wireframes and align development with patient context. We then compared the designs against the personas and storyboards, using heuristic evaluations and conducted usability testing with real patients to check workflows and reduce friction before the pilot study with the developed platform.

\section{Pilot Study and Engagement Findings}

A mixed-methods pilot study was conducted over two months in Summer and Fall 2020 with seven participants reporting medical complexity. The pilot aimed to solicit feedback on the developed platform in participants’ real-world daily lives and determine readiness for larger future studies. Participants with varied self-reported chronic conditions participated in intake sessions to document their care journeys and challenges, followed by platform use in daily routines, and concluded with follow-up sessions for feedback. Engagement was evaluated using a combination of objective usage analytics (e.g., login frequency and feature utilization) to measure adoption patterns and subjective self-reports of perceived usefulness. This multimodal evaluation aimed to integrate behavioral data with insights into capacity elements like cognitive and emotional resources, providing a comprehensive view of engagement under real-world conditions.

Participants did not use the platform frequently, and several described being too busy, having no capacity, and being occupied with other responsibilities for their care. This pilot took place during the COVID-19 pandemic, which could have introduced additional contextual stresses. 
Participants' engagement breakdowns were described as the platform creating more work rather than support, pointing to challenges like too many steps to click through and the inability to backdate or edit entries. Further issues were seen in symptom tracking constraints and a lack of clarity around measurement units. We reflect on these engagement patterns through three lessons about why engagement was difficult to sustain: (1) the inherent challenges of MVP simplicity failing to meet viability thresholds for high-complexity patients, (2) process-level drift during implementation that erodes initial patient-centered priorities, and (3) cross-condition interpretive burden can undermine continued use when patients must manually connect fragmented information across conditions.

\section{Lessons Learned for Sustaining Engagement:}

 \subsection{Technical and patient-oriented minimally viable product are not the same when designing for engagement under patient complexity:}
A lesson from this project is that offering a minimally viable product (MVP) to patients can fail to support engagement. 
For patients living with medical complexity who have variable, multiple chronic conditions and existing treatment burden, added value must notably reduce workload, increase capacity, or do both to be viable and adopted. In our pilot, the implemented version was technically functional, but over-simplification made it difficult to integrate into everyday life. For example, tracking did not allow patients to backdate or fix data entry errors, and planned insight visualizations were not included in the implemented version. This technical simplification offloaded additional complexity and stress onto patients who were already overwhelmed, helping explain why continued use was hard to justify under workload and capacity constraints. This lesson raises research questions:
\begin{itemize}
\item How can early versions deliver enough value and a strong first impression that justifies continued engagement after the first try?

\end{itemize}

\subsection{The importance of maintaining patient-centered intent through different phases of implementation in cross-stakeholder projects:}
While using a HCD process 
allowed us to capture patient needs and capacity-sensitive supports, the implemented DHI did not preserve those priorities in the pilot. This can be difficult to carry patient-centered intent through implementation when clinical realities and software delivery constraints must be balanced. This occurred because the project spanned multiple disciplines, from medicine (clinical realities), computing (software delivery constraints), and patient perspectives, requiring constant balancing of different priorities. Under feasibility, budget, and timeline pressures, “simplifications” can unintentionally shift workload back onto patients unless teams use explicit alignment mechanisms to keep patient-centered outcomes visible and make tradeoffs legible as scope changes. This suggests that engagement breakdown can arise when design intent and implementation decisions drift out of alignment, not only from interaction details in the final product. A question can be discussed in the workshop: 
\begin{itemize}

    \item What practices help teams maintain alignment on patient-centered engagement priorities as projects move from design intent into implementation decisions? 

\end{itemize}

 \subsection{Siloed features can undermine engagement by offloading interpretive work across conditions:}
Without automation for cross-condition integration (e.g., third-party data pulls, pattern detection linking heart rate fluctuations to multiple comorbidities), the platform required patients to manually connect and interpret isolated metrics, quietly shifting cognitive sensemaking labor onto people. 
This perpetuated the "messenger" role described by patients, where tools rarely supported holistic, boundary-crossing insights. Pilot feedback highlighted how this design choice forced extra interpretive effort that participants could not spare amid multimorbidity and stress. As a result, the added cognitive and coordination load exacerbated workload-capacity imbalances (per CuCoM), making the platform feel like additional work rather than meaningful support, contributing to infrequent use and disengagement in the pilot.
Questions include:
\begin{itemize}
\item What are effective ways to measure the hidden cognitive and sensemaking demands that contribute to disengagement in patient-facing tools for complex chronic care?
\item How can early-stage DHIs (MVPs or prototypes) strike a balance between technical minimalism and sufficient automation to make the tool feel "worth trying"?
\item How can we define and measure the threshold for interpretive/sensemaking effort in DHIs for complex chronic care, where exceeding it shifts the tool from supportive to burdensome and drives disengagement?

\end{itemize}

These lessons point to a simple but demanding requirement for engagement under complexity: early tools must be worth trying and worth returning to, without asking patients to take on additional coordination or interpretation work. Our pilot highlights how easily engagement can falter when viability thresholds are missed, when patient-centered priorities are difficult to preserve through implementation, and when cross-condition sensemaking is left to the user. We hope this contribution helps the workshop surface shared language, methods, and evaluation approaches for building capacity-sensitive tools that can be taken up and sustained in everyday life.

\section{Acknowledgments}
This work is supported in part by NSERC Discovery Grant: Interactive Visualization RGPIN-2019-07192, and Canada Research Chair in Data Visualization CRC-2019-00368.

\bibliographystyle{plainnat}
\bibliography{acmart}

\end{document}